\documentclass{aa}

\usepackage{graphicx}
\usepackage{txfonts}
\usepackage{color}
\usepackage{stfloats}
\usepackage{amsmath}
\usepackage{multirow}

\usepackage{epsfig}
\usepackage{amsmath}
\usepackage{subfigure}
\usepackage{natbib}

\usepackage{longtable}
\usepackage{graphicx}
\usepackage{epstopdf}
\usepackage{booktabs}
\usepackage{txfonts}
\usepackage[utf8]{inputenc}%
\usepackage{graphicx}
\usepackage{txfonts}
\usepackage{stfloats}

\newcommand{\tablenotea}[1]{\parbox{9.0cm}{\indent \footnotesize{#1}}}
\newcommand{\tablenoteb}[1]{\parbox{18.3cm}{\indent \footnotesize{#1}}}

\newcommand{\acp}{Adv. Chem. Phys.}

\newcommand{\cpl}{Chem. Phys. Lett.}
\newcommand{\chemrev}{Chem. Rev.}

\newcommand{\ijqc}{Int. J. Quantum Chem.}

\newcommand{\jms}{J. Mol. Spectr.}
\newcommand{\jmst}{J. Mol. Struct.}
\newcommand{\joc}{J. Organomet. Chem.}
\newcommand{\jpc}{J. Phys. Chem.}
\newcommand{\jpca}{J. Phys. Chem. A}

\newcommand{\nature}{Nature}

\newcommand{\nps}{Nature Phys. Sci.}

\newcommand{\prev}{Phys. Rev.}

\begin{document}

\title{Interstellar detection of the simplest aminocarbyne, H$_2$NC:\\ an ignored but abundant molecule\thanks{Based on observations carried out with the IRAM 30m Telescope. IRAM is supported by INSU/CNRS (France), MPG (Germany) and IGN (Spain). $\dag$ These authors contributed equally to this work.}}

\titlerunning{Interstellar H$_2$NC}
\authorrunning{Cabezas et al.}

\author{C.~Cabezas\inst{1}$^{\dag}$, M.~Ag\'undez\inst{1}$^{\dag}$, N.~Marcelino\inst{1}, B.~Tercero\inst{2,3}, S.~Cuadrado\inst{1}, \and J.~Cernicharo\inst{1}}

\institute{
Instituto de F\'isica Fundamental, CSIC, C/ Serrano 123, 28006 Madrid, Spain\\
\email{carlos.cabezas@csic.es, marcelino.agundez@csic.es} \and
Observatorio Astron\'omico Nacional, IGN, Calle Alfonso XII 3, E-28014 Madrid, Spain \and
Observatorio de Yebes, IGN, Cerro de la Palera s/n, E-19141 Yebes, Guadalajara, Spain
}

\date{Received; accepted}


\abstract
{We report the first identification in space of H$_2$NC, a high-energy isomer of H$_2$CN that has been largely ignored in chemical and astrochemical studies. The observation of various unidentified lines around 72.2\,GHz in the cold dark cloud L483 motivated the search for, and successful detection of, additional groups of lines in harmonic relation. After an exhaustive high-level \textit{ab initio} screening of possible carriers, we confidently assign the unidentified lines to H$_2$NC based on the good agreement between astronomical and theoretical spectroscopic parameters and sound spectroscopic and astrochemical arguments. The observed frequencies are used to precisely characterize the rotational spectrum of H$_2$NC. This species is also detected in the cold dark cloud B1-b and the $z$\,=\,0.89 galaxy in front of the quasar PKS\,1830$-$211. We derive H$_2$NC/H$_2$CN abundance ratios $\sim$\,1 in L483 and B1-b and 0.27 toward PKS\,1830$-$211. Neither H$_2$NC nor H$_2$CN are detected in the dark cloud TMC-1, which seriously questions a previous identification of H$_2$CN in this source. We suggest that the H$_2$NC/H$_2$CN ratio behaves as the HNC/HCN ratio, with values close to one in cold dense clouds and below one in diffuse clouds. The reactions N + CH$_3$ and C + NH$_3$ emerge as strong candidates to produce H$_2$NC in interstellar clouds. Further studies on these two reactions are needed to evaluate the yield of H$_2$NC. Due to the small number of atoms involved, it should be feasible to constrain the chemistry behind H$_2$NC and H$_2$CN, just as it has been done for HNC and HCN, as this could allow to use the H$_2$NC/H$_2$CN ratio as a probe of chemical or physical conditions of the host clouds.}


\maketitle

\section{Introduction}

High spectral resolution line surveys of molecular clouds are invaluable tools to do molecular spectroscopy in space and to study fundamental chemical processes. Since the discovery of HCO$^+$, first assigned to an unidentified line \citep{Buhl1970} based on \textit{ab initio} calculations \citep{Wahlgren1973} and later on confirmed in the laboratory \citep{Woods1975}, several molecular species have been detected in space prior to their characterization in the laboratory, in most cases aided by state-of-the-art \textit{ab initio} calculations. Examples of this kind are HCS$^+$ \citep{Thaddeus1981}, C$_4$H \citep{Guelin1978}, C$_6$H \citep{Suzuki1986}, and C$_3$H$^+$ \citep{Pety2012}, all them confirmed later on in the laboratory, as well as the cases of C$_5$N$^-$ \citep{Cernicharo2008}, MgC$_3$N and MgC$_4$H \citep{Cernicharo2019}, and HC$_5$NH$^+$ \citep{Marcelino2020}, which have not been yet observed in the laboratory. If the source displays narrow lines and spectral resolution is high enough, it is possible to resolve the hyperfine structure and derive very precise spectroscopic parameters (e.g., \citealt{Cernicharo1987}).

Here we present a new case of molecular spectroscopy done in space. Thanks to sensitive high-spectral resolution observations of the cold dark cloud L483 at mm wavelengths we have detected various groups of unidentified lines, revealing a complex hyperfine structure, which we confidently assign to H$_2$NC. This species is the simplest member of the family of aminocarbyne ligands, well known in organometallic chemistry \citep{Pombeiro2001}. It is also a high-energy metastable isomer of the methylene amidogen radical (H$_2$CN), and as such it has been largely ignored in many chemical and astrochemical studies. We however find that in the cold dark clouds L483 and B1-b, H$_2$NC has the same abundance than H$_2$CN itself. These results illustrate nicely how the chemical composition of cold interstellar clouds is driven by chemical kinetics rather than by thermochemical considerations, unlike suggested by \cite{Lattelais2009}. We discuss which chemical reactions could be behind the formation of H$_2$NC, something that has been overlooked in the literature.

\section{Astronomical observations}

\begin{figure*}
\centering
\includegraphics[angle=0,width=\textwidth]{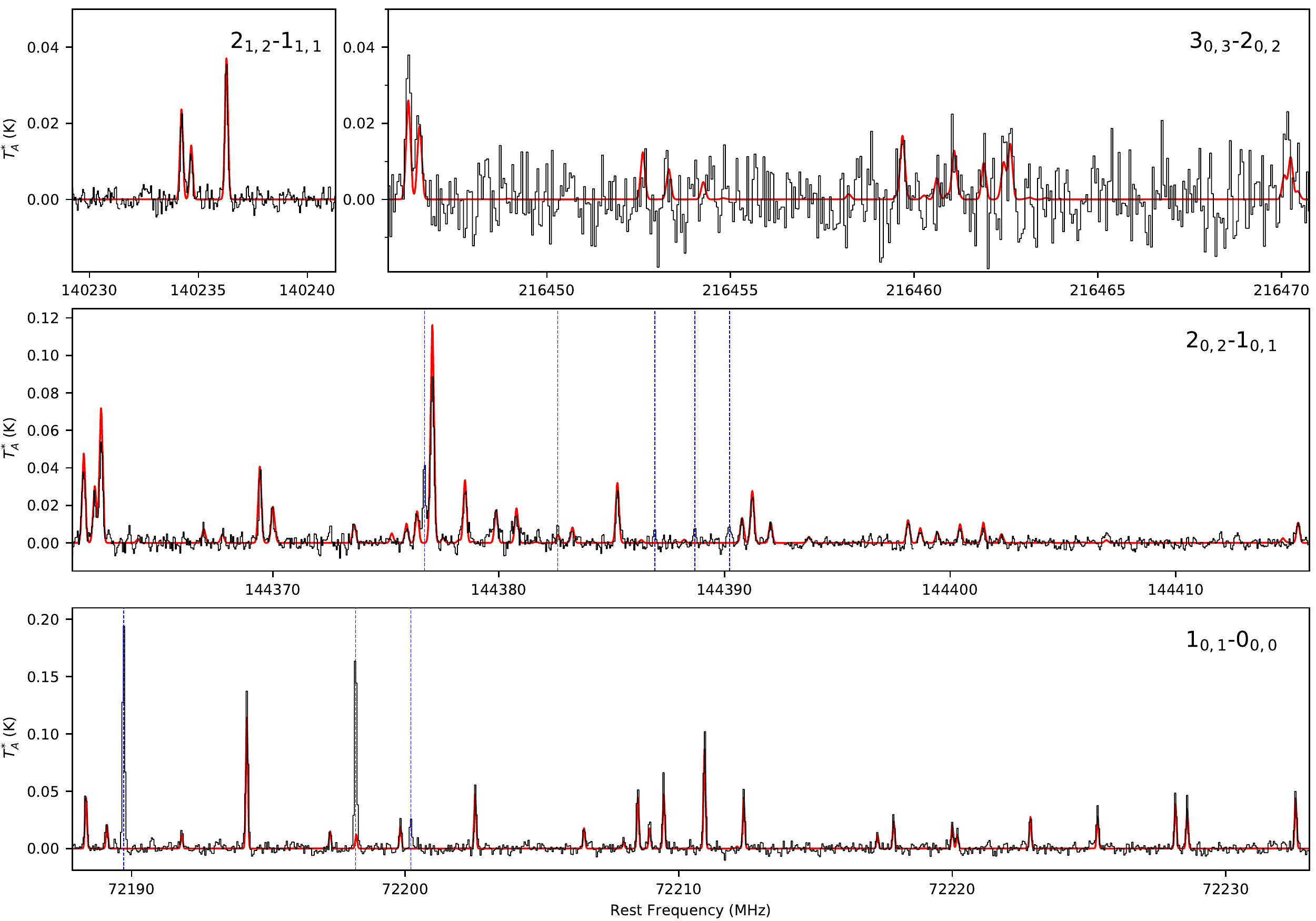}
\caption{Lines of H$_2$NC observed in L483 (see parameters in Table~\ref{table:lines}). In red computed synthetic spectra for a column density of H$_2$NC of 10$^{12}$\,cm$^{-2}$, an ortho-to-para ratio of 3, a rotational temperature of 4.0\,K, a full width at half maximum of 0.35 km s$^{-1}$ (except for 3$_{0,3}$-2$_{0,2}$ lines, in which case a value of 0.19 km s$^{-1}$ is adopted), and an emission size that fills the IRAM\,30m main beam. Vertical blue lines indicate the position of CCD lines (see Appendix~\ref{sec:ccd}).} \label{fig:h2nc_l483}
\end{figure*}

We recently carried out a line survey of the cold dark cloud L483 in the 80-116\,GHz frequency range with the IRAM\,30m telescope \citep{Agundez2019}. Additional observations of L483 were carried out in December 2018 to cover the 72-80\,GHz range. A group of lines spread over a frequency interval of 45\,MHz around 72.2\,GHz, close to the hyperfine components of the $N$\,=\,1-0 transition of CCD, caught our attention, as we could not assign them using either the {\small CDMS}\footnote{\tiny{\texttt{https://cdms.astro.uni-koeln.de}}} \citep{Muller2005}, {\small JPL}\footnote{\tiny{\texttt{https://spec.jpl.nasa.gov}}} \citep{Pickett1998}, or {\small MADEX}\footnote{\tiny{\texttt{https://nanocosmos.iff.csic.es/madex}}} \citep{Cernicharo2012_madex} catalogues. The spectral pattern is typical of a molecule with a complex fine and hyperfine structure. Since the hyperfine structure usually collapses for increasing rotational quantum number, the large splitting observed suggests that the lines correspond to a rotational transition with a low rotational quantum number. Our first hypothesis was that the set of unidentified lines corresponds to the $N$\,=\,1-0 rotational transition of a linear or quasi-linear species. The rotational constant $B$, or ($B$\,+\,$C$)/2, would then be 36.1\,GHz, which implies that the species must contain two heavy atoms (C, N, O) and possibly one or various H atoms. If the lines corresponded to a higher rotational transition, e.g., $N$\,=\,2-1, the immediately higher transition, e.g., $N$\,=\,3-2, would fall below 116\,GHz, and we have no hint of it in our sensitive $\lambda$\,3\,mm data. Moreover, we are confident that the lines do not arise from a heavy isotopologue involving D, $^{13}$C, $^{15}$N, $^{17}$O, or $^{18}$O because this would imply to have a similar pattern of lines from the parent species at higher frequencies and significantly more intense (see the case of \mbox{HDCCN}; \citealt{Cabezas2021}), and there is no feature in our $\lambda$\,3\,mm consistent with this possibility. The only species that fulfill these criteria, such as HCN, C$_2$H, and N$_2$H$^+$, have the rotational spectrum of their minor isotopologues well characterized in the laboratory.

\begin{table*}
\small
\caption{Line parameters of H$_2$NC observed in L483.}
\label{table:lines}
\centering
\begin{tabular}{{c@{\hspace{0.38cm}}c@{\hspace{0.38cm}}c@{\hspace{0.38cm}}c@{\hspace{0.38cm}}c@{\hspace{0.38cm}}cc@{\hspace{0.38cm}}c@{\hspace{0.38cm}}c@{\hspace{0.38cm}}c@{\hspace{0.38cm}}c@{\hspace{0.38cm}}crrccc}}
\hline \hline
 $N'$ & $K'_a$ & $K'_c$ & $J'$ & $F'_1$ & $F'$ & $N''$ & $K''_a$ & $K''_c$ & $J''$ & $F''_1$ & $F''$ & \multicolumn{1}{c}{$\nu_{obs}$} & $\nu_{o}-\nu_{c}$ & \multicolumn{1}{c}{$\Delta v$} & \multicolumn{1}{c}{$T_A^*$ peak} & \multicolumn{1}{c}{$\int T_A^* dv$} \\
 &  &   &   &   &  &  &   &  &   &  &   & \multicolumn{1}{c}{(MHz)}  & \multicolumn{1}{c}{(MHz)} & \multicolumn{1}{c}{(km s$^{-1}$)} & \multicolumn{1}{c}{(mK)} & \multicolumn{1}{c}{(mK km s$^{-1}$)} \\
\hline
1 & 0 & 1 & 1.5 & 1.5 & 2.5 & 0 & 0 & 0 & 0.5 & 0.5 & 1.5 &  72188.328(10) & $-$0.007    & 0.41(3)  &  52 & 22.5(11) \\
1 & 0 & 1 & 1.5 & 0.5 & 1.5 & 0 & 0 & 0 & 0.5 & 0.5 & 0.5 &  72189.077(10) & $-$0.024    & 0.45(6)  &  21 & 10.1(10) \\
1 & 0 & 1 & 1.5 & 2.5 & 1.5 & 0 & 0 & 0 & 0.5 & 1.5 & 1.5 &  72191.838(10) &  0.006    & 0.40(12) &  17 &  7.3(15) \\
1 & 0 & 1 & 1.5 & 2.5 & 3.5 & 0 & 0 & 0 & 0.5 & 1.5 & 2.5 &  72194.219(10) &  0.008    & 0.38(1)  & 137 & 56.0(12) \\
1 & 0 & 1 & 0.5 & 0.5 & 1.5 & 0 & 0 & 0 & 0.5 & 0.5 & 0.5 &  72197.262(10) &  0.005    & 0.20(10) &  26 &  5.6(14) \\
1 & 0 & 1 & 1.5 & 0.5 & 1.5 & 0 & 0 & 0 & 0.5 & 0.5 & 1.5 &  72199.831(10) &  0.003    & 0.37(4)  &  27 & 10.4(10) \\
1 & 0 & 1 & 1.5 & 2.5 & 1.5 & 0 & 0 & 0 & 0.5 & 1.5 & 0.5 &  72202.571(10) &  0.012    & 0.35(2)  &  57 & 21.2(11) \\
1 & 0 & 1 & 1.5 & 0.5 & 0.5 & 0 & 0 & 0 & 0.5 & 1.5 & 1.5 &  72206.545(10) &  0.006    & 0.46(11) &  17 &  8.5(13) \\
1 & 0 & 1 & 1.5 & 1.5 & 2.5 & 0 & 0 & 0 & 0.5 & 1.5 & 2.5 &  72208.502(10) & $-$0.005    & 0.37(3)  &  56 & 22.3(10) \\
1 & 0 & 1 & 1.5 & 1.5 & 0.5 & 0 & 0 & 0 & 0.5 & 1.5 & 0.5 &  72208.937(10) & $-$0.002    & 0.41(7)  &  26 & 11.3(11) \\
1 & 0 & 1 & 1.5 & 1.5 & 1.5 & 0 & 0 & 0 & 0.5 & 1.5 & 1.5 &  72209.450(10) & $-$0.004    & 0.35(2)  &  66 & 24.7(10) \\
1 & 0 & 1 & 1.5 & 2.5 & 2.5 & 0 & 0 & 0 & 0.5 & 1.5 & 1.5 &  72210.947(10) &  0.007    & 0.38(1)  & 106 & 42.4(12) \\
1 & 0 & 1 & 0.5 & 1.5 & 2.5 & 0 & 0 & 0 & 0.5 & 0.5 & 1.5 &  72212.383(10) &  0.007    & 0.33(2)  &  53 & 18.2(10) \\
1 & 0 & 1 & 1.5 & 0.5 & 0.5 & 0 & 0 & 0 & 0.5 & 1.5 & 0.5 &  72217.255(10) & $-$0.010    & 0.35(8)  &  14 &  5.3(15) \\
1 & 0 & 1 & 0.5 & 1.5 & 1.5 & 0 & 0 & 0 & 0.5 & 0.5 & 0.5 &  72217.855(10) & $-$0.002    & 0.33(4)  &  32 & 11.1(10) \\
1 & 0 & 1 & 1.5 & 0.5 & 1.5 & 0 & 0 & 0 & 0.5 & 1.5 & 2.5 &  72219.996(10) & $-$0.005    & 0.35(5)  &  23 &  8.6(11) \\
1 & 0 & 1 & 1.5 & 1.5 & 1.5 & 0 & 0 & 0 & 0.5 & 1.5 & 0.5 &  72220.178(10) & $-$0.003    & 0.36(8)  &  18 &  6.8(12) \\
1 & 0 & 1 & 0.5 & 0.5 & 0.5 & 0 & 0 & 0 & 0.5 & 0.5 & 1.5 &  72222.854(10) & $-$0.001    & 0.36(4)  &  30 & 11.7(10) \\
1 & 0 & 1 & 0.5 & 1.5 & 0.5 & 0 & 0 & 0 & 0.5 & 0.5 & 0.5 &  72225.310(10) & $-$0.003    & 0.44(4)  &  36 & 17.2(11) \\
1 & 0 & 1 & 0.5 & 0.5 & 1.5 & 0 & 0 & 0 & 0.5 & 1.5 & 2.5 &  72228.155(10) & $-$0.001    & 0.36(3)  &  50 & 19.1(12) \\
1 & 0 & 1 & 0.5 & 1.5 & 1.5 & 0 & 0 & 0 & 0.5 & 0.5 & 1.5 &  72228.583(10) & $-$0.002    & 0.32(3)  &  46 & 15.9(9)  \\
1 & 0 & 1 & 0.5 & 1.5 & 2.5 & 0 & 0 & 0 & 0.5 & 1.5 & 2.5 &  72232.548(10) & $-$0.001    & 0.37(3)  &  50 & 19.8(11) \\
2 & 1 & 2 & 2.5 & 2.5 &     & 1 & 1 & 1 & 1.5 & 1.5 &     & 140234.242(10) &  0.024    & 0.35(3)  &  22 &  8.2(10) \\
2 & 1 & 2 & 2.5 & 1.5 &     & 1 & 1 & 1 & 1.5 & 0.5 &     & 140234.652(10) & $-$0.015    & 0.36(6)  &  12 &  4.5(11) \\
2 & 1 & 2 & 2.5 & 3.5 &     & 1 & 1 & 1 & 1.5 & 2.5 &     & 140236.279(10) & $-$0.002    & 0.31(2)  &  37 & 12.3(10) \\
2 & 0 & 2 & 2.5 & 2.5 & 3.5 & 1 & 0 & 1 & 1.5 & 1.5 & 2.5 & 144361.610(10) & $-$0.004    & 0.38(3)  &  38 & 15.6(9)  \\
2 & 0 & 2 & 2.5 & 1.5 & 2.5 & 1 & 0 & 1 & 1.5 & 0.5 & 1.5 & 144362.101(10) & $-$0.003    & 0.43(8)  &  23 & 10.5(10) \\
2 & 0 & 2 & 2.5 & 3.5 & 4.5 & 1 & 0 & 1 & 1.5 & 2.5 & 3.5 & 144362.392(10) &  0.010    & 0.35(3)  &  54 & 19.9(11) \\
2 & 0 & 2 & 2.5 & 1.5 & 0.5 & 1 & 0 & 1 & 1.5 & 0.5 & 0.5 & 144366.915(10) & $-$0.020    & 0.15(3)  &  11 &  1.8(6)  \\
2 & 0 & 2 & 1.5 & 0.5 & 1.5 & 1 & 0 & 1 & 0.5 & 0.5 & 0.5 & 144367.795(15) &  0.040    & 0.22(7)  &   8 &  1.7(5)  \\
2 & 0 & 2 & 2.5 & 3.5 & 2.5 & 1 & 0 & 1 & 1.5 & 2.5 & 1.5 & 144369.440(10) &  0.024    & 0.27(2)  &  40 & 11.4(10) \\
2 & 0 & 2 & 2.5 & 2.5 & 1.5 & 1 & 0 & 1 & 1.5 & 1.5 & 0.5 & 144369.970(10) & $-$0.014    & 0.35(5)  &  18 &  7.0(11) \\
2 & 0 & 2 & 2.5 & 1.5 & 2.5 & 1 & 0 & 1 & 1.5 & 1.5 & 2.5 & 144373.607(15) &  0.010    & 0.33(5)  &  11 &  3.7(8)  \\
2 & 0 & 2 & 2.5 & 2.5 & 3.5 & 1 & 0 & 1 & 1.5 & 2.5 & 3.5 & 144375.925(15) &  0.016    & 0.36(7)  &   8 &  3.0(7)  \\
2 & 0 & 2 & 2.5 & 2.5 & 1.5 & 1 & 0 & 1 & 1.5 & 2.5 & 1.5 & \multirow{2}{*}{\bigg\} 144376.393(10)} & \multirow{2}{*}{0.013} & \multirow{2}{*}{0.45(4)}  &  \multirow{2}{*}{16} &  \multirow{2}{*}{7.7(11)} \\ 
2 & 0 & 2 & 1.5 & 1.5 & 2.5 & 1 & 0 & 1 & 0.5 & 1.5 & 2.5 &                                         &                    &                           &                       &  \\
2 & 0 & 2 & 2.5 & 3.5 & 3.5 & 1 & 0 & 1 & 1.5 & 2.5 & 2.5 & \multirow{2}{*}{\bigg\} 144377.058(10)} & \multirow{2}{*}{$-$0.008}     & \multirow{2}{*}{0.39(1)}  &  \multirow{2}{*}{85} & \multirow{2}{*}{35.8(13)} \\ 
2 & 0 & 2 & 1.5 & 2.5 & 3.5 & 1 & 0 & 1 & 0.5 & 1.5 & 2.5 &                           &      &               &       & \\
2 & 0 & 2 & 2.5 & 2.5 & 2.5 & 1 & 0 & 1 & 1.5 & 1.5 & 1.5 & 144378.508(10) &  0.004  & 0.56(6)  &  24 & 14.4(10) \\
2 & 0 & 2 & 2.5 & 1.5 & 1.5 & 1 & 0 & 1 & 1.5 & 0.5 & 0.5 & 144379.862(15) & $-$0.020  & 0.46(10) &  17 &  8.2(10) \\
2 & 0 & 2 & 1.5 & 1.5 & 2.5 & 1 & 0 & 1 & 0.5 & 0.5 & 1.5 & 144380.762(15) & $-$0.022  & 0.40(8)  &  14 &  6.1(9)  \\
2 & 0 & 2 & 1.5 & 1.5 & 1.5 & 1 & 0 & 1 & 0.5 & 0.5 & 0.5 & 144383.214(20) & $-$0.049  & 0.29(9)  &   7 &  2.1(8)  \\
2 & 0 & 2 & 1.5 & 2.5 & 2.5 & 1 & 0 & 1 & 0.5 & 1.5 & 1.5 & 144385.268(10) &  0.010  & 0.34(3)  &  26 &  9.5(9)  \\
2 & 0 & 2 & 1.5 & 0.5 & 1.5 & 1 & 0 & 1 & 1.5 & 0.5 & 1.5 & 144390.759(10) & $-$0.022  & 0.31(5)  &  13 &  4.2(9)  \\
2 & 0 & 2 & 1.5 & 2.5 & 1.5 & 1 & 0 & 1 & 0.5 & 1.5 & 0.5 & \multirow{2}{*}{\bigg\} 144391.233(10)} & \multirow{2}{*}{$-$0.002} & \multirow{2}{*}{0.36(3)}  &  \multirow{2}{*}{25} &  \multirow{2}{*}{9.6(11)} \\
2 & 0 & 2 & 1.5 & 0.5 & 0.5 & 1 & 0 & 1 & 0.5 & 0.5 & 1.5 &                           &      &               &       &  \\
2 & 0 & 2 & 1.5 & 1.5 & 0.5 & 1 & 0 & 1 & 0.5 & 0.5 & 0.5 & 144392.039(10) & $-$0.005  & 0.28(6)  &  10 &  3.2(8)  \\
2 & 0 & 2 & 1.5 & 1.5 & 1.5 & 1 & 0 & 1 & 0.5 & 0.5 & 1.5 & 144398.146(10) &  0.014  & 0.31(3)  &  12 &  3.8(7)  \\
2 & 0 & 2 & 1.5 & 2.5 & 1.5 & 1 & 0 & 1 & 0.5 & 1.5 & 1.5 & 144398.681(20) &  0.007  & 0.38(7)  &   6 &  2.5(6)  \\
2 & 0 & 2 & 1.5 & 0.5 & 0.5 & 1 & 0 & 1 & 1.5 & 0.5 & 1.5 & 144399.434(15) &  0.006  & 0.20(10) &   7 &  1.4(4)  \\
2 & 0 & 2 & 1.5 & 1.5 & 2.5 & 1 & 0 & 1 & 1.5 & 1.5 & 2.5 & 144400.428(15) & $-$0.005  & 0.32(6)  &   8 &  2.7(6)  \\
2 & 0 & 2 & 1.5 & 2.5 & 2.5 & 1 & 0 & 1 & 0.5 & 1.5 & 2.5 & 144401.458(15) & $-$0.008  & 0.32(7)  &   7 &  2.5(5)  \\
2 & 0 & 2 & 1.5 & 2.5 & 3.5 & 1 & 0 & 1 & 1.5 & 2.5 & 3.5 & 144415.423(10) &  0.008  & 0.37(6)  &  11 &  4.2(6)  \\
3 & 0 & 3 & 3.5 & 3.5 & 4.5 & 2 & 0 & 2 & 2.5 & 2.5 & 3.5 & \multirow{2}{*}{\bigg\}216446.230(15)} &   \multirow{2}{*}{0.003}   & \multirow{2}{*}{0.20(3)}  &  \multirow{2}{*}{40} &  \multirow{2}{*}{8.3(7)}  \\ 
3 & 0 & 3 & 3.5 & 2.5 & 3.5 & 2 & 0 & 2 & 2.5 & 1.5 & 2.5 &                           &      &               &       & \\
3 & 0 & 3 & 3.5 & 4.5 & 5.5 & 2 & 0 & 2 & 2.5 & 3.5 & 4.5 & 216446.512(15) & $-$0.021     & 0.27(5)  &  22 &  6.4(8)  \\
3 & 0 & 3 & 3.5 & 4.5 & 4.5 & 2 & 0 & 2 & 2.5 & 3.5 & 3.5 & 216459.686(20) &  0.010     & 0.16(6)  &  18 &  3.0(6)  \\
3 & 0 & 3 & 3.5 & 2.5 & 2.5 & 2 & 0 & 2 & 2.5 & 1.5 & 1.5 & 216462.452(20) &  0.018     & 0.14(4)  &  19 &  2.9(6)  \\
3 & 0 & 3 & 2.5 & 3.5 & 4.5 & 2 & 0 & 2 & 1.5 & 2.5 & 3.5 & 216462.639(20) &  0.028     & 0.15(6)  &  21 &  3.4(7)  \\
3 & 0 & 3 & 2.5 & 2.5 & 2.5 & 2 & 0 & 2 & 1.5 & 1.5 & 1.5 & \multirow{2}{*}{\bigg\}216470.144(20)} &   \multirow{2}{*}{$-$0.037}   & \multirow{2}{*}{0.22(5)}  &  \multirow{2}{*}{22} &  \multirow{2}{*}{5.8(2)}  \\ 
3 & 0 & 3 & 2.5 & 3.5 & 3.5 & 2 & 0 & 2 & 1.5 & 2.5 & 2.5 &                           &      &               &       & \\
\hline
\end{tabular}
\tablenoteb{Numbers in parentheses are 1$\sigma$ uncertainties in units of the last digits. Line parameters were derived from a Gaussian fit to the line profile. Observed frequencies are derived adopting $V_{\rm LSR}$ = 5.30 km s$^{-1}$ \citep{Agundez2019}. Errors in the observed frequencies are estimated from the Gaussian fit and range between 10 kHz, for the lines detected with high signal-to-noise ratios, and 20 kHz, for the weakest lines. $\nu_{o}-\nu_{c}$ is the observed minus calculated frequency, where the calculated frequency comes from the fit described in Sect.~\ref{sec:assignment}. $\Delta v$ is the full width at half maximum.}
\end{table*}

In order to confirm our hypothesis we requested IRAM\,30m telescope time to search for the two immediately higher transitions. Observations were carried out in June 2019 and we detected unidentified lines at the expected positions of the $N$\,=\,2-1 and $N$\,=\,3-2 transitions, 144\,GHz and 216\,GHz, respectively, thus confirming our initial hypothesis. Moreover, we also detected unidentified lines around 140\,GHz, which were later on assigned to the para species of H$_2$NC. These observations, and the ones performed before, were done using the frequency-switching technique with a frequency throw of 7.2\,MHz. To be sure that no line is missing due to the potential cancellation with negative artifacts, at $\pm$\,7.2\,MHz of each line, caused by this procedure, we carried out new observations around 72.2\,GHz, 140\,GHz, 144\,GHz, and 216\,GHz in November 2019, June 2020, and February and April 2021, using the wobbler-switching technique, with the secondary mirror nutating by 220$''$ at a rate of 0.5\,Hz. All observations were done using the EMIR receivers connected to a fast Fourier transform spectrometer providing a spectral resolution of 50\,kHz \citep{Klein2012}. Line intensities are expressed in terms of the antenna temperature, $T_A^*$, which can be converted to main beam brightness temperature, $T_{\rm mb}$, by dividing by $B_{\rm eff}$/$F_{\rm eff}$. IRAM\,30m beam efficiencies, forward efficiencies, and half power beam widths are given on the IRAM website\footnote{\tiny{\texttt{http://www.iram.es/IRAMES/mainWiki/Iram30mEfficiencies}}}. Further details on the IRAM\,30m observations of L483 can be found in \cite{Agundez2019} and \cite{Agundez2021}. The observed lines are shown in Fig.~\ref{fig:h2nc_l483} and the frequencies and line parameters derived are given in Table~\ref{table:lines}.

We note that the frequencies of some of the hyperfine components of the $N$\,=\,1-0 transition of CCD observed in L483 differ significantly from those given in the {\small CDMS} catalogue \citep{Muller2005}. This was already noticed by \cite{Yoshida2019} in their observations of L1527. We therefore used the new frequencies to improve the spectroscopic parameters of CCD (see Appendix~\ref{sec:ccd}).

We also make use of observations of the cold dark clouds B1-b and TMC-1, and of the $z$\,=\,0.89 galaxy in front of the quasar PKS\,1830$-$211. The observations of B1-b and TMC-1 were carried out with the IRAM\,30m telescope in the 72-74\,GHz range in August 2018 and are part of a $\lambda$\,3\,mm line survey of these two sources (see \citealt{Cernicharo2012} for more details), while those of PKS\,1830$-$211 were done using the Yebes\,40m telescope in the Q band in a monitoring campaign of this source from April 2019 to July 2020 (see \citealt{Tercero2020} for more details). Here we added data taken during 8 observing sessions from October 2019 to the data presented in \cite{Tercero2020}. All data were reduced using the program {\small CLASS} of the {\small GILDAS} software\footnote{\tiny{\texttt{http://www.iram.fr/IRAMFR/GILDAS}}}.

\section{Spectroscopic assignment to H$_2$NC} \label{sec:assignment}

Potential carriers of the unidentified lines observed in L483 are N$_2$H$_x$, CNH$_x$, COH$_x$, NOH$_x$, C$_2$H$_x$, and O$_2$H$_x$, either neutral or ionic. To obtain precise spectroscopic parameters that help in the assignment of the observed lines, we carried out high-level \textit{ab initio} calculations for all plausible candidates. The geometry optimization calculations for all the candidates considered were done using the spin-restricted coupled cluster method with single, double, and perturbative triple excitations (RCCSD(T); \citealt{Raghavachari1989}), with all electrons (valence and core) correlated and the Dunning's correlation consistent basis sets with polarized core-valence correlation quadruple-$\zeta$ (cc-pCVQZ; \citealt{Woon1995}). These calculations were carried out using the Molpro 2020.2 program \citep{Werner2020}. In addition to the rotational constants, we calculated other parameters necessary to interpret the rotational spectrum. At the optimized geometries of each species we calculated the fine and hyperfine constants. They are the three spin-rotation coupling constants ($\varepsilon_{aa}$, $\varepsilon_{bb}$, and $\varepsilon_{cc}$), the magnetic hyperfine constants for the non zero nuclear spin nuclei ($a_F$, $T_{aa}$, and $T_{bb}$), and the nuclear electric quadrupole constants for the nitrogen nucleus ($\chi_{aa}$ and $\chi_{bb}$), if the molecule contains any. The spin-rotation coupling constants were calculated using the second order M{\o}ller-Plesset perturbation (MP2; \citealt{Moller1934}) with the aug-cc-pVQZ basis set. On the other hand, the other hyperfine constants were calculated at the quadratic configuration interaction with single and double excitations (QCISD; \citealt{Pople1987}) level of calculation with the aug-cc-pVQZ basis set. Harmonic frequencies were computed at UCCSD/aug-cc-pVQZ level of theory \citep{Cizek1969} to estimate the centrifugal distortion constants. These calculations were done using the Gaussian16 program \citep{Frisch2016}. Table~\ref{table:abinitio} summarizes the results of these calculations.

Given the observed splitting of $\sim$45\,MHz, it is very unlikely that the carrier is a closed-shell species. Even if the molecule has two N nuclei, the hyperfine splitting would be of a few MHz and the number of lines would be smaller than the observed ones. Hence, the carrier should be an open-shell species containing at least two nuclei with a non zero nuclear spin, so that rotational transitions have fine and hyperfine structure. Taking this into account, closed-shell species such as $cis$-HNNH, H$_2$NN, H$_2$CN$^-$, H$_2$NC$^+$, H$_2$NO$^+$, $cis$-HNOH$^+$, and $trans$-HNOH$^+$ are discarded. The value of ($B$+$C$)/2 derived from astronomical observations is 36.1\,GHz, and thus the rotational constants calculated for NNH, HCN$^+$, HCN$^-$, H$_2$CN$^+$, $cis$-HCNH$^+$, $trans$-HCNH$^+$, and H$_2$NC$^+$ clearly exclude them as carriers. In the cases of NNH$^-$, H$_2$NN$^+$, $trans$-HCNH, and $cis$-HCNH, the calculated ($B$+$C$)/2 are closer to the astronomical value, although the differences are larger than 2.5\,\%. These four candidates can be discarded because the level of calculation used provides errors not larger than 1\,\%. In addition, their predicted splitting does not agree with the observed one. For NNH$^-$, H$_2$NN$^+$, and $trans$-HCNH, the hyperfine components of the 1$_{0,1}$-0$_{0,0}$ transition span over 100-200\,MHz, while for $cis$-HCNH, the splitting is very small, around 10\,MHz. At this point, only three species show a ($B$+$C$)/2 value compatible with the observed one, 36.1\,GHz. H$_2$NC has the closest ($B$+$C$)/2 value, while $cis$-HCOH$^+$ and $trans$-HCOH$^+$ have values a bit larger, but also compatible taking into account the accuracy of the calculations. However, $cis$-HCOH$^+$ and $trans$-HCOH$^+$ have only two nuclei with non zero nuclear spin, and thus, the number of hyperfine components and their splitting pattern cannot explain all the unidentified lines observed in L483. In contrast, H$_2$NC has three nuclei with non zero nuclear spin and its predicted splitting is consistent with the observed one. To obtain accurate spectroscopic parameters for our best candidate, H$_2$NC, we scaled the calculated values using experimental/theoretical ratios derived for the isoelectronic species H$_2$CN, for which experimental rotational parameters are known \citep{Yamamoto1992}. This procedure has been found to provide rotational constants with an accuracy better than 0.1\,\% (e.g., \citealt{Cabezas2021}). The theoretical rotational constants $A$, $B$, and $C$ obtained by this way are given in Table~\ref{table:constants}. It is seen that the calculated ($B$+$C$)/2, 36.092\,GHz, is very similar to the observed value, which prompted us to consider H$_2$NC as the starting point in our data analysis.

\begin{table}
\small
\caption{Calculated molecular data for the candidates considered.}
\label{table:abinitio}
\centering
\begin{tabular}{l@{\hspace{0.2cm}}c@{\hspace{0.2cm}}c@{\hspace{0.8cm}}l@{\hspace{0.2cm}}c@{\hspace{0.2cm}}c}
\hline
\hline
Species & ($B$+$C$)/2 & G.S. & Species & ($B$+$C$)/2 & G.S.  \\
\hline
NNH          & 44.968       & $^{2}A'$     & H$_2$CN$^+$      & 31.626 & $^{3}A_2$ \\
NNH$^-$      & 37.605       & $^{3}A''$    & H$_2$CN$^-$      & 35.117 & $^{1}A_1$ \\
$cis$-HNNH   & 36.763       & $^{1}A'$     & $cis$-HCNH$^+$   & 34.271 & $^{3}A'$  \\
H$_2$NN      & 36.558       & $^{1}A'$     & $trans$-HCNH$^+$ & 33.004 & $^{3}A''$ \\
H$_2$NN$^+$  & 37.971       & $^{2}B_2$    & H$_2$NC$^+$      & 38.369 & $^{1}A_1$ \\
HCN$^+$      & 40.073\,$^a$ & $^{2}\Pi$    & $cis$-HCOH$^+$   & 36.654 & $^{2}A'$  \\
HCN$^-$      & 44.432\,$^a$ & $^{2}\Sigma$ & $trans$-HCOH$^+$ & 36.534 & $^{2}A'$  \\
$trans$-HCNH & 37.125       & $^{2}A'$     & H$_2$NO$^+$      & 36.815 & $^{1}A_1$ \\
$cis$-HCNH   & 37.330       & $^{2}A'$     & $cis$-HNOH$^+$   & 35.312 & $^{1}A'$  \\
H$_2$NC      & 36.214       & $^{2}B_2$    & $trans$-HNOH$^+$ & 35.591 & $^{1}A'$  \\
\hline
\end{tabular}
\tablefoot{($B$+$C$)/2 in units of GHz. G.S. stands for ground electronic state. $^a$\,Value corresponds to $B$ because this species is linear.}
\end{table}

The observed frequencies (see Table~\ref{table:lines}) were analyzed with the SPFIT program \citep{Pickett1991} using an appropriate Hamiltonian for an asymmetric top with a doublet electronic state ($^2B_2$) and C$_{2v}$ symmetry. The employed Hamiltonian has the following form:
\begin{equation}
\textbf{H} = \textbf{H}_{rot} + \textbf{H}_{sr} + \textbf{H}_{mhf} + \textbf{H}_Q
\end{equation}
where \textbf{H}$_{rot}$ contains rotational and centrifugal distortion parameters, $\textbf{H}_{sr}$ is the spin-rotation term, $\textbf{H}_{mhf}$ represents the magnetic hyperfine coupling interaction term due to the N and H nuclei, and $\textbf{H}_Q$ represent the nuclear electric quadrupole interaction due to the N nucleus. The coupling scheme used is \textbf{J}\,=\,\textbf{N}\,+\,\textbf{S}, \textbf{F}$_1$\,=\,\textbf{J}\,+\,\textbf{I}$_1$, and \textbf{F}\,=\,\textbf{F}$_1$\,+\,\textbf{I}$_2$, where \textbf{I}$_1$\,=\,\textbf{I}(N) and \textbf{I}$_2$\,=\,\textbf{I}(H$_1$)\,+\,\textbf{I}(H$_2$). H$_2$NC has two equivalent H nuclei and it is thus necessary to discern between $ortho$ and $para$ species. The $ortho$ levels are described by $K_a$ + $K_c$ even while the $para$ levels by $K_a$ + $K_c$ odd. The hyperfine interaction term \textbf{H}$_{mhf}$ is thus written explicitly as a two spin system:
\begin{equation}
\textbf{H}_{mhf} = a_F^{\rm(N)} \cdot \textbf{S} \cdot \textbf{I}_1 + \textbf{I}_1 \cdot \textbf{T}^{\rm(N)} \cdot \textbf{S} + a_F^{\rm(H_1,H_2)} \cdot \textbf{S} \cdot \textbf{I}_2 + \textbf{I}_2 \cdot \textbf{T}^{\rm(H_1,H_2)} \cdot \textbf{S}
\end{equation}
where $a_F^{\rm(N)}$ and \textbf{T}$^{\rm(N)}$ stand for the Fermi contact constant and the dipole-dipole interaction tensor for the N nucleus, respectively, and $a_F^{\rm(H_1,H_2)}$ and \textbf{T}$^{\rm(H_1,H_2)}$ are averages of the coupling constants for the two H nuclei. There is a term proportional to the difference of the coupling constants for the dipole-dipole interaction. However, the term connects levels only off–diagonal in $K_a$, and thus it can be ignored, treating the two H nuclei as if they are equivalent. In this manner, each energy level is denoted by six quantum numbers: $N$, $K_a$, $K_c$, $J$, $F_1$, and $F$.

The results obtained from the fit are shown in Table~\ref{table:constants}. We experimentally determined the values for a total of thirteen molecular parameters, considering the two equivalent H nuclei. These parameters are the rotational constants $B$ and $C$, the centrifugal distortion constant $\Delta_N$, the electron spin-rotation coupling constants $\varepsilon_{bb}$ and $\varepsilon_{cc}$, the magnetic hyperfine constants for the nitrogen and hydrogen nuclei $a_F$ and $T_{aa}$, and the nuclear electric quadrupole constant for the nitrogen nucleus $\chi_{aa}$. Other constants were kept fixed to the theoretical values calculated in this work. The standard deviation of the fit is 15.2\,kHz.

\begin{table}
\small
\caption{Spectroscopic parameters of H$_2$NC (all in MHz).}
\label{table:constants}
\centering
\begin{tabular}{lcc}
\hline \hline
\multicolumn{1}{c}{Parameter}  & \multicolumn{1}{c}{Astronomical (L483) fit\,$^a$} & \multicolumn{1}{c}{Theoretical\,$^b$} \\
\hline
$A$                       &  ~    339669.82         &  ~  339669.82     \\
$B$                       &  ~  38085.559(74)       &  ~   38102.01     \\
$C$                       &  ~    34124.301(74)     &  ~   34082.34     \\
$\Delta_N$                &  ~     0.13463(16)      &  ~      0.101     \\
$\Delta_{NK}$             &  ~        2.367         &  ~      2.367     \\
$\Delta_K$                &  ~        26.907        &  ~     26.907     \\
$\delta_N$                &  ~        0.012         &  ~      0.012     \\
$\delta_K$                &  ~        1.758         &  ~      1.758     \\
$\varepsilon_{aa}$        &  ~      2319.573        &  ~   2319.573     \\
$\varepsilon_{bb}$        &  ~       1.88(59)       &  ~      9.221     \\
$\varepsilon_{cc}$        &  ~    $-$57.17(58)      &  ~  $-$54.056     \\
$a_F$$^{\rm(N)}$          &  ~   $-$22.8656(59)     &  ~  $-$21.248     \\
$T_{aa}$$^{\rm(N)}$       &  ~      3.417(14)       &  ~      3.464     \\
$T_{bb}$$^{\rm(N)}$       &  ~        0.581         &  ~      0.581     \\
$\chi_{aa}$$^{\rm(N)}$    &  ~      0.2125(83)      &  ~      0.212     \\
$\chi_{bb}$$^{\rm(N)}$    &  ~       1.594       &  ~      1.594     \\
$a_F$ $^{\rm(H_1,H_2)}$   &  ~      169.90(23)      &  ~    170.360     \\
$T_{aa}$$^{\rm(H_1,H_2)}$ &  ~      9.345(13)       &  ~      8.648     \\
$T_{bb}$$^{\rm(H_1,H_2)}$ &  ~     $-$2.675         &  ~   $-$2.675     \\
\hline
$|\mu|$                 &                           &  ~   3.83\,$^c$     \\

\hline
\end{tabular}
\tablenotea{$^a$\,Numbers in parentheses are 1$\sigma$ uncertainties in units of the last digits. Parameters without uncertainties were fixed to the theoretical values. $^b$\,The values of the rotational constants $A$, $B$, and $C$ were corrected using the experimental and theoretical constants of H$_2$CN. $^c$\,Calculated at the RCCSD(T)/cc-pCVQZ level of theory, in units of Debye.}
\end{table}

\begin{figure}
\centering
\includegraphics[angle=0,width=0.90\columnwidth]{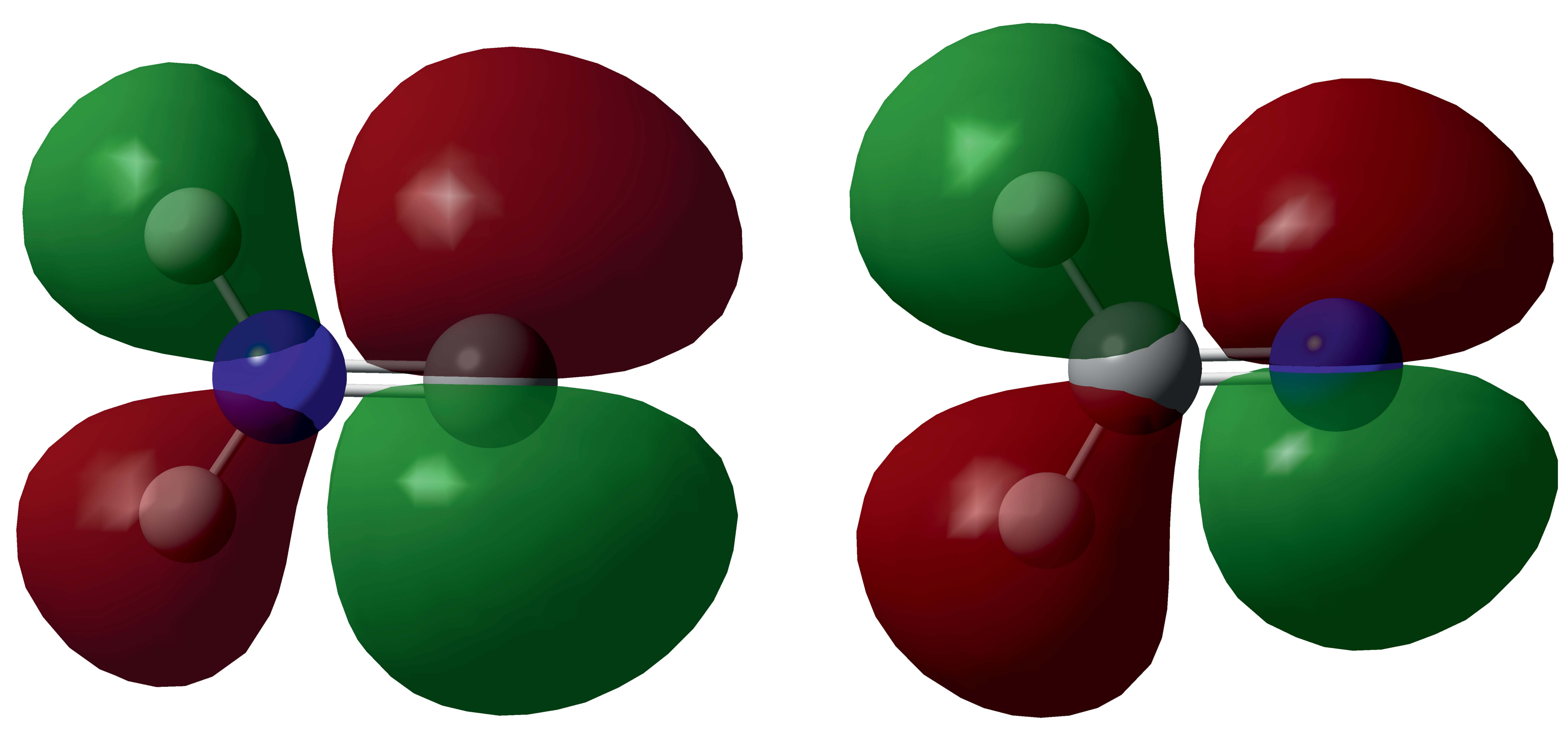}
\caption{Geometries and molecular orbital for the unpaired electron in H$_2$NC (left) and H$_2$CN (right).} \label{fermi}
\end{figure}

The results in Table~\ref{table:constants} show an excellent agreement between the theoretical and astronomical values of the rotational constants $B$ and $C$. The relative errors are 0.01\,\% and 0.06\,\% for $B$ and $C$, respectively. These errors are slightly larger if we compare with the unscaled theoretical values of $B$ and $C$, 0.04\,\% and 0.58\,\%, respectively. The rotational constants provide information on the mass distribution of the molecular species and, thus, the good accordance between observed and calculated values could serve by itself to identify the spectral carrier. Moreover, the high resolution of our survey allows us to completely scrutinize the fine and hyperfine structure of the rotational transitions. This structure depends strongly on the electronic environment of the molecular species, which is related to the number of nuclei with a non zero nuclear spin and their relative positions. Hence, the fine and hyperfine coupling constants determined are ideal fingerprints of the spectral carrier, providing key information for its identification. The results from our fit indicate that our carrier must have two equivalent hydrogen nuclei and one nitrogen nucleus, whose $a_F$ and $T_{aa}$ constants are in very good agreement with those predicted for H$_2$NC. The $a_F$ and $T_{aa}$ values for the two H nuclei are comparable to those of H$_2$CN, 233.152\,MHz and 8.294\,MHz, respectively \citep{Yamamoto1992}, which is reasonable since their electronic environment is almost similar in both molecules. However, the $a_F$ and $T_{aa}$ for the N nucleus are very different in H$_2$NC and H$_2$CN. First, the $T_{aa}$ values for H$_2$NC and H$_2$CN are 3.413\,MHz and $-$45.143\,MHz, respectively, which indicates that the spin density on the N nucleus is much smaller in H$_2$NC. Secondly, the Fermi constants $a_F$ have opposite signs, $-$22.8660\,MHz for H$_2$NC and 25.916 MHz for H$_2$CN. This is explained by the different mechanisms for spin polarization due to the unpaired electron on both systems. Figure~\ref{fermi} shows the molecular orbital of the unpaired electron in the two species. In the case of H$_2$NC, the orbital has a node at the location of the N atom, while for H$_2$CN the N atom is completely covered by this orbital. The rest of parameters determined from our fit, $\Delta_N$, $\varepsilon_{bb}$, and $\varepsilon_{cc}$, agree reasonably well with the values calculated theoretically for H$_2$NC. However, it should be mentioned that the accuracy of the prediction of the constants $\varepsilon_{aa}$, $\varepsilon_{bb}$, and $\varepsilon_{cc}$ is usually much smaller than for the rotational or hyperfine constants. In the light of all these arguments we definitively conclude that the carrier of the unidentified lines observed in L483 is H$_2$NC.

\section{Discussion} \label{sec:discussion}

\subsection{The H$_2$NC/H$_2$CN ratio in the interstellar medium}

We computed synthetic spectra assuming local thermodynamic equilibrium to derive the column density of H$_2$NC in L483. The relative intensities of the various rotational transitions observed constrain the rotational temperature to a low value, 4.0\,$\pm$\,0.2\,K, which is consistent with a highly polar carrier like H$_2$NC. The calculated dipole moment of H$_2$NC is 3.83\,D (see Table~\ref{table:constants}). As full width at half maximum we adopt the average of the observed values, 0.19 km s$^{-1}$ for the 3$_{0,3}$-2$_{0,2}$ lines and 0.35 km s$^{-1}$ for the rest (see Table~\ref{table:lines}). We derive a beam-averaged column density of (1.0\,$\pm$\,0.2)\,$\times$\,10$^{12}$ cm$^{-2}$ for H$_2$NC in L483 (see synthetic spectrum in Fig.~\ref{fig:h2nc_l483}). The ortho-to-para ratio is fully consistent with the statistical value of three.

H$_2$NC is the highest energy member of the isomeric family composed of H$_2$CN, $trans$-HCNH, $cis$-HCNH, and H$_2$NC. Our calculations predict that H$_2$CN is the most stable isomer, with $trans$-HCNH and $cis$-HCNH lying above by 7.9 kcal mol$^{-1}$ and 12.8 kcal mol$^{-1}$, respectively, while H$_2$NC lies 29.9 kcal mol$^{-1}$ higher in energy with respect to H$_2$CN. These results are in agreement with those reported by \citet{Puzzarini2010}.

\begin{figure}
\centering
\includegraphics[angle=0,width=\columnwidth]{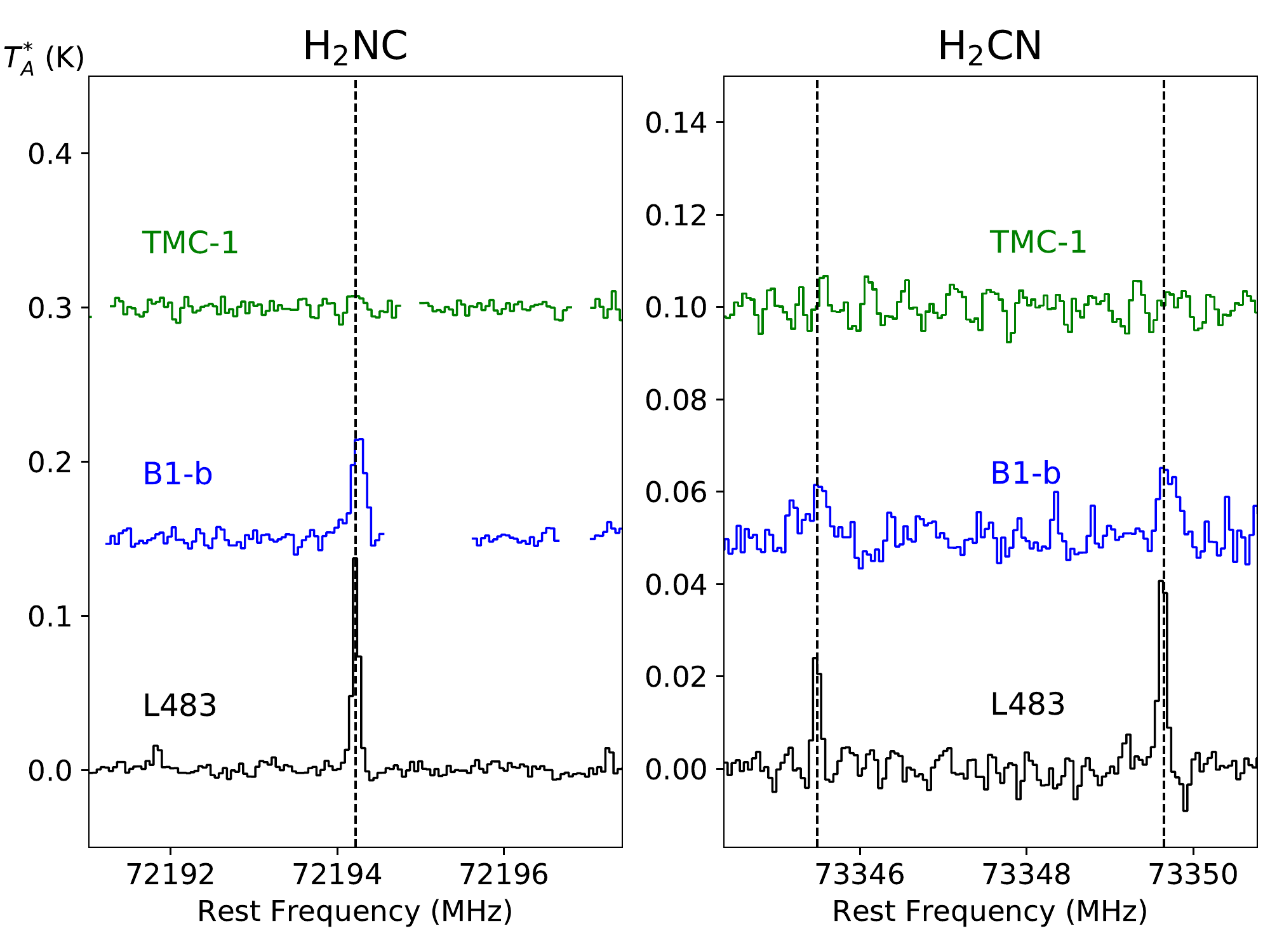}
\caption{Strongest hyperfine components of H$_2$NC and H$_2$CN in L483, B1-b, and TMC-1. None of the two isomers are detected in TMC-1.} \label{fig:sources}
\end{figure}

A further argument in support of the assignment to H$_2$NC comes from astronomical observations, which indicate that the carrier is related to H$_2$CN. This radical is also observed in L483 (see \citealt{Agundez2019} and Fig.~\ref{fig:sources}). From the observed intensities of the 1$_{0,1}$-0$_{0,0}$ and 2$_{0,2}$-1$_{0,1}$ lines of H$_2$CN in L483 we derive a rotational temperature of 4 K and a column density of (8\,$\pm$\,2)\,$\times$10$^{11}$ cm$^{-2}$. This value is somewhat lower than that given by \cite{Agundez2019} because at that time line strengths were incorrectly computed downward by a factor of three and now we also include the 2$_{0,2}$-1$_{0,1}$ transition. The H$_2$NC/H$_2$CN abundance ratio in L483 is thus 1.25\,$\pm$\,0.30. Our carrier, H$_2$NC, is also observed in B1-b, where H$_2$CN is detected as well. In B1-b we derive $N$(H$_2$NC)\,=\,(1.0\,$\pm$\,0.2)\,$\times$\,10$^{12}$ cm$^{-2}$ and $N$(H$_2$CN)\,=(1.0\,$\pm$\,0.2)\,$\times$\,10$^{12}$ cm$^{-2}$, which results in a H$_2$NC/H$_2$CN abundance ratio of 1.0\,$\pm$\,0.3, similar to the value found in L483. Moreover, in the well known cold dark cloud TMC-1, which harbors a rich variety of molecules, neither H$_2$NC nor H$_2$CN are detected in our data. We derive 3$\sigma$ upper limits to their column densities of 3.2\,$\times$10$^{11}$ cm$^{-2}$ and 4.8\,$\times$\,10$^{11}$ cm$^{-2}$, respectively. \cite{Ohishi1994} reported the discovery of H$_2$CN in TMC-1. Although the column density derived by these authors, 1.5\,$\times$\,10$^{11}$ cm$^{-2}$, is consistent with our 3$\sigma$ upper limit, the identification of \cite{Ohishi1994} is questionable because the two lines detected are only marginally seen above their noise level and they should result in antenna temperatures above 20 mK at IRAM\,30m, which is not consistent with our data (see Fig.~\ref{fig:sources}). Therefore, both our carrier and H$_2$CN are seen in L483 and B1-b with a similar abundance ratio, while none of them are seen in TMC-1. This fact strengthens the idea that our carrier is related to H$_2$CN.

\begin{figure}
\centering
\includegraphics[angle=0,width=\columnwidth]{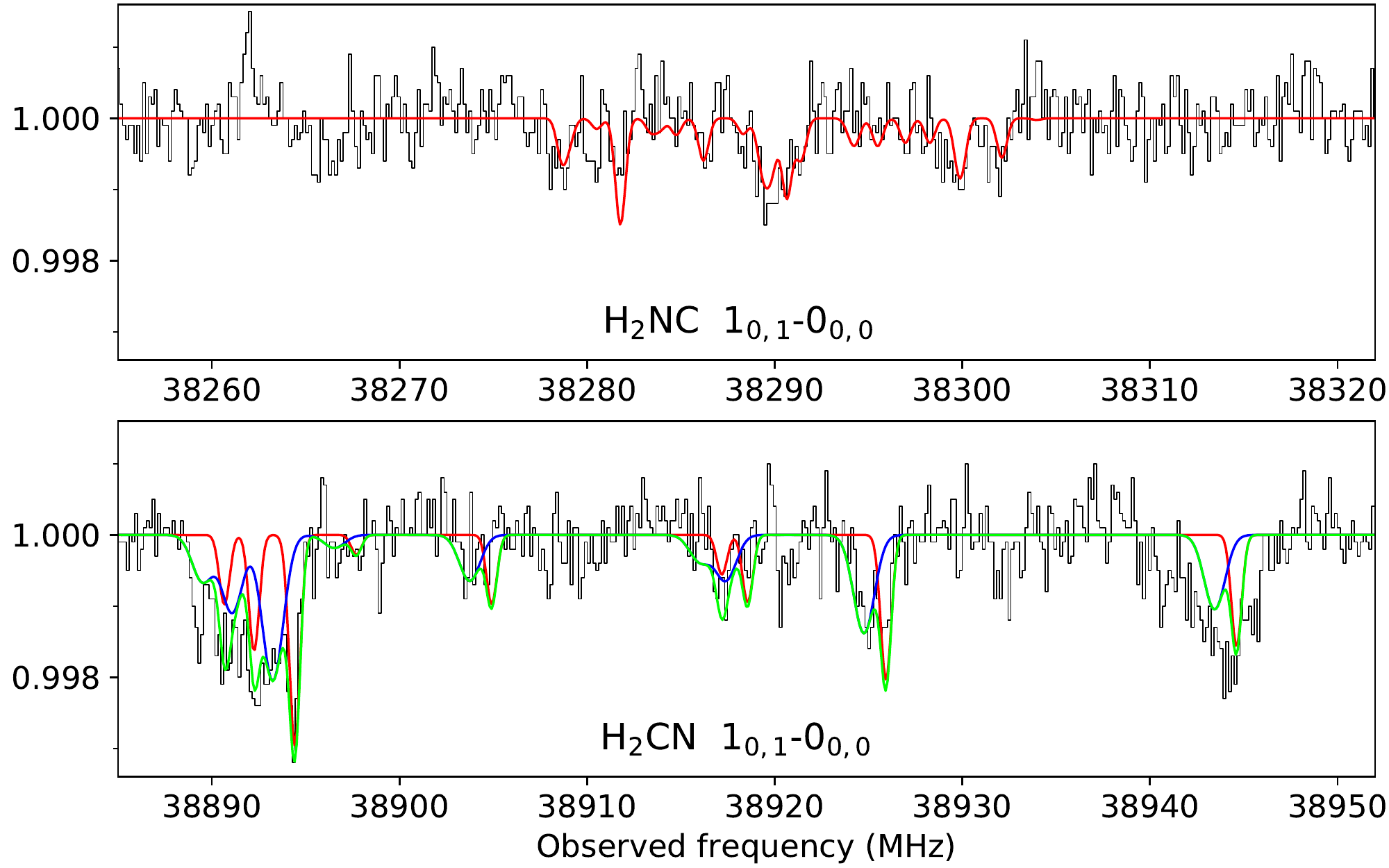}
\caption{Absorption lines of H$_2$NC and H$_2$CN observed with the Yebes\,40m telescope towards PKS\,1830$-$211. We assume a redshift of $z$\,=\,0.885875 to compute the rest frequency. The $y$-axis is the intensity normalized to the total continuum level. The red lines are the synthetic spectra computed for one component, while in the H$_2$CN spectrum, the blue line corresponds to an additional component and the green line to the sum of the two components. See more details in \cite{Tercero2020}.} \label{fig:pks}
\end{figure}

The fact that the strongest hyperfine component of H$_2$NC is 3-4 times more intense that the strongest one of H$_2$CN in L483 and B1-b suggests that H$_2$NC could be easily found in those sources where H$_2$CN is detected. H$_2$CN has been reported toward L1544 \citep{Vastel2019} and the $z$\,=\,0.89 galaxy in front of the quasar PKS\,1830$-$211 \citep{Tercero2020}. We examined our Yebes\,40m data of PKS\,1830$-$211 and we find significant absorption at the position of the strongest hyperfine components of H$_2$NC 1$_{0,1}$-0$_{0,0}$ (see Fig.~\ref{fig:pks}). We computed synthetic spectra for H$_2$NC and H$_2$CN (red curves in Fig.~\ref{fig:pks}) assuming the same physical parameters of the source reported by \cite{Tercero2020}. We derive a H$_2$NC/H$_2$CN abundance ratio of 0.27\,$\pm$\,0.08. The new data added to the PKS\,1830$-$211 spectrum allow us to clearly distinguish an additional component for H$_2$CN (blue curve; see \citealt{Tercero2020}. The fact that the H$_2$NC/H$_2$CN ratio is significantly lower than the values found in L483 and B1-b points to a chemical differentiation between this high-redshift galaxy and galactic cold interstellar clouds. In fact, the chemical composition in this lensing galaxy is more characteristic of diffuse or translucent clouds rather than of cold dense clouds, as indicated by the presence of ions such as C$_3$H$^+$ \citep{Tercero2020}. The case of H$_2$NC and H$_2$CN isomers reminds that of HNC and HCN. The HNC/HCN abundance ratio is $\sim$\,1 in cold dense clouds \citep{Hirota1998,Sarrasin2010} and $<$\,1 in diffuse clouds \citep{Liszt2001}. In the gravitational lens of PKS\,1830$-$211 the abundance ratio HNC/HCN is 0.4 \citep{Muller2006}, in line with expectations if the source has a diffuse cloud character. Thus, similarly to the case of HNC and HCN, it is likely that in general cold dense clouds show H$_2$NC/H$_2$CN ratios around one and diffuse clouds H$_2$NC/H$_2$CN ratios below one. The presence of H$_2$NC is therefore a further example of how chemical kinetics, rather than thermodynamics, regulate the chemical composition of interstellar clouds.

\subsection{The chemistry of H$_2$NC and H$_2$CN}

The isomer H$_2$NC is missing in chemical kinetics databases used in astrochemistry, such as {\small UMIST} \citep{McElroy2013} or {\small KIDA} \citep{Wakelam2015}. However, the more stable isomer H$_2$CN is included and it is thus interesting to look at its main formation reactions. To this purpose, we run a pseudo time-dependent gas-phase chemical model with typical parameters of cold dark clouds (see, e.g., \citealt{Agundez2013}). We used the chemical network {\small RATE12} from the {\small UMIST} database \citep{McElroy2013}, with updates relevant for the chemistry of H$_2$CN from \cite{Loison2015} and \cite{Hickson2015}. According to our calculations, H$_2$CN is formed with a peak abundance relative to H$_2$ of $\sim$\,10$^{-10}$. The main formation reactions are C + NH$_3$, N + CH$_3$, and N + CH$_2$CN, while it is mainly destroyed through reactions with neutral H, C, N, and O atoms. Taking into account that H$_2$NC and H$_2$CN are observed with similar abundances in L483 and B1-b, it is likely that both isomers share a common formation route.

\begin{figure*}
\centering
\includegraphics[angle=0,width=\textwidth]{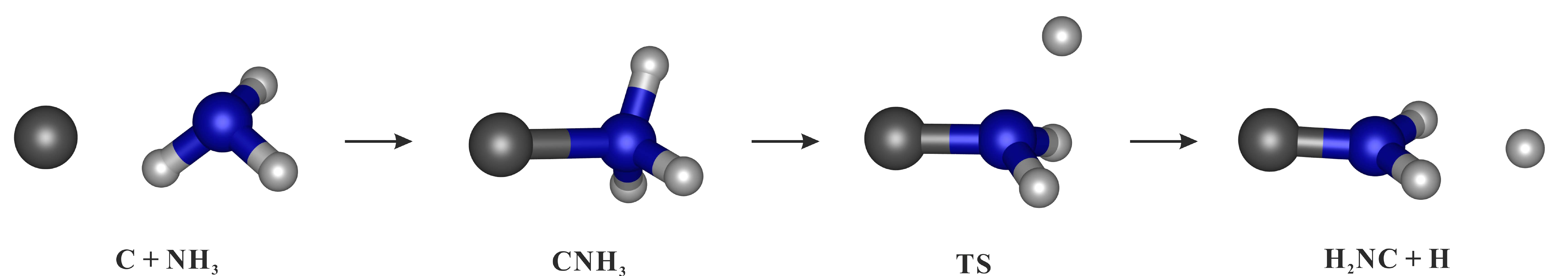}
\caption{Molecular structures of the species involved in the formation of H$_2$NC in the C + NH$_3$ reaction. TS stands for transition state.} \label{fig:c_nh3_reaction}
\end{figure*}

The reaction between N and CH$_3$ has been studied experimentally and theoretically. Experimental studies in the temperature range 200-423\,K reveal a rate coefficient around 10$^{-10}$ cm$^3$ s$^{-1}$, with a positive temperature dependence, and a branching ratio of 0.9 for production of H atoms \citep{Marston1989a,Marston1989b}. Computational studies point to H$_2$CN + H as main channel \citep{Cimas2006,Alves2008}. The study of \cite{Alves2008} does not consider H$_2$NC as product, but that of \cite{Cimas2006} considers H$_2$NC and $trans$-HCNH. \cite{Cimas2006} show how the reactants N and CH$_3$ approach through an attractive potential surface leading to an intermediate, H$_3$CN, whose formation does not involve any barrier. This intermediate has two different possibilities for its further evolution. In the first place, elimination of a hydrogen atom would lead to H$_2$CN + H, the most exothermic channel. This process involves a transition state which lies well below the reactants. A second possibility is isomerization into  H$_2$CNH, through hydrogen migration from carbon to nitrogen, which proceeds through a transition state also located well below the reactants. H$_2$CNH may lead to $trans$-HCNH + H involving a lower barrier and a further isomerization into HCNH$_2$ would produce the H$_2$NC + H channel as final product. Although the three isomers, H$_2$CN, $trans$-HCNH and H$_2$NC can be formed without barrier (all the transition states are submerged below the reactants), elimination of a hydrogen atom from the initially formed intermediate, H$_3$CN, leading to H$_2$CN + H is the preferred reaction path. \cite{Cimas2006} state that residual quantities of $trans$-HCNH can be formed, whereas for H$_2$NC the branching ratio is negligible.

The reaction C + NH$_3$ has been studied more recently, and it has been found that it is rapid at 50\,K, with a rate coefficient of 1.8\,$\times$\,10$^{-10}$ cm$^3$ s$^{-1}$ \citep{Hickson2015}. In a related study, \cite{Bourgalais2015} combined photoionization and LIF experiments supported by theoretical calculations and showed that the H$_2$CN + H channel represents 100\% of the product yield for this reaction. These authors rule out that $trans$-HCNH or $cis$-HCNH are formed, but do not discuss whether H$_2$NC can be produced. The theoretical calculations reported by \cite{Bourgalais2015} show that the first step is the formation of a reaction intermediate, CNH$_3$, in a process without any barrier. As in the case of the H$_3$CN intermediate in the N + CH$_3$ reaction discussed above, CNH$_3$ can proceed through two different pathways. The first one, the elimination of a hydrogen atom, would lead to H$_2$NC + H, through a transition state which lies below the reactants. This reaction path is illustrated in Fig.~\ref{fig:c_nh3_reaction}. On the other hand, the isomerization of the CNH$_3$ intermediate through hydrogen migration from nitrogen to carbon can lead to HCNH$_2$. This isomerization proceeds through a transition state also located below the reactants. HCNH$_2$ can further evolve to form H$_2$NC + H or $trans$-HNCH. A second hydrogen migration from nitrogen to carbon in HCNH$_2$ produces H$_2$CNH, which ultimately leads H$_2$CN + H, that is the most exothermic channel. Although H$_2$NC + H is not the most exothermic product channel, it could be the most favored channel if we consider the elimination of a hydrogen atom from CNH$_3$ as the preferred reaction path, in analogy to the N + CH$_3$ results of \cite{Cimas2006}. 

Another fact that reinforces the previous statement is that the identification of H$_2$CN as main product in the C + NH$_3$ reaction by \cite{Bourgalais2015} is based exclusively in the agreement between their measured photoionization spectrum and that measured and attributed to H$_2$CN by \cite{Nesbitt1991}. However, as discussed by \citet{Holzmeier2013}, the ionization energy measured by \cite{Nesbitt1991} is derived from an ion yield curve and it is therefore afflicted with a rather large uncertainty. In fact, \citet{Holzmeier2013} measured an excitation energy of 12.32 eV for H$_2$CN, very different from the value derived by \cite{Nesbitt1991}, 10.8 eV. This can be attributed to the fact that the method used by \cite{Nesbitt1991} contains less information than the threshold photoelectron spectra of \citet{Holzmeier2013}, which precluded \cite{Nesbitt1991} from distinguishing between isomers. Since the experimental methods employed by \cite{Bourgalais2015} and \cite{Nesbitt1991} are quite similar, it is reasonable to think that both have the same limitations. Hence, if H$_2$NC is formed in the experiment of \cite{Bourgalais2015}, it could have escaped detection with their experimental setup.

The reaction between N and CH$_2$CN was considered by \cite{Loison2015} to yield H$_2$CN with a moderately high rate coefficient by analogy to the reaction N + CH$_3$. However, no dedicated studies have been carried out on this particular reaction and thus it is uncertain whether H$_2$CN or any of its metastable isomers are formed. Other plausible routes to H$_2$CN and H$_2$NC are the reactions between radicals NH + CH$_2$ and NH$_2$ + CH, although they are probably less efficient than N + CH$_3$ and C + NH$_3$ because the reactants are expected to have lower abundances. Formation routes involving ions are also possible, in particular those leading to the various possible isomers H$_x$CNH$_{3-x}$$^+$ ($x$\,=\,0-3), which upon dissociative recombination with electrons can form H$_2$CN and H$_2$NC. A plausible such route could involve the reaction NH$_4^+$ + C, although its rate coefficient and products are not known.

In summary, the two isomers H$_2$CN and H$_2$NC can be produced in a barrierless process in the reactions N + CH$_3$ and C + NH$_3$, with H$_2$NC being a more likely product of the latter reaction. In the light of our results, it would be interesting to revisit these two reactions and evaluate the yields of the different isomers H$_2$CN, $trans$- and $cis$-HCNH, and H$_2$NC. Other routes are also possible, although theoretical or experimental information is missing.

\section{Conclusions}

We observed various sets of unidentified lines in the cold dark cloud L483, which we confidently assign to H$_2$NC, a high-energy metastable isomer of H$_2$CN. The astronomical lines are used to precisely characterize the rotational spectrum of H$_2$NC. Both H$_2$NC and H$_2$CN are detected, in addition to L483, in the cold dark cloud B1-b and the high redshift lensing galaxy in front of the quasar PKS\,1830$-$211. Neither H$_2$NC nor H$_2$CN are detected in the dark cloud TMC-1, which puts in question a previous claim of detection of H$_2$CN in this source. We derive H$_2$NC/H$_2$CN abundance ratios $\sim$\,1 in the cold dense clouds L483 and B1-b and 0.27 in the high redshift galaxy. It is suggested that the H$_2$NC/H$_2$CN ratio behaves as the HNC/HCN ratio, with values around one in cold dense clouds and below one in diffuse and translucent media. The most obvious formation routes to H$_2$CN and H$_2$NC are the reactions N + CH$_3$ and C + NH$_3$. The latter is particularly favorable for production of H$_2$NC. Further studies on these reactions, able to constrain the yield of each isomer, should allow to reach a good understanding of the chemistry behind H$_2$NC and H$_2$CN. This could open the door to use the H$_2$NC/H$_2$CN ratio as a proxy of chemical or physical parameters of interstellar clouds.

\begin{acknowledgements}

We acknowledge funding support from Spanish Ministerio de Ciencia e Innovaci\'on through grants AYA2016-75066-C2-1-P, PID2019-106110GB-I00, and PID2019-107115GB-C21, and from the European Research Council (ERC Grant 610256: NANOCOSMOS). M.A. also acknowledges funding support from the Ram\'on y Cajal programme of Spanish Ministerio de Ciencia e Innovaci\'on (grant RyC-2014-16277). We thank Prof. Ingo Fischer for his comments on the photoionization spectrum of the methylene amidogen radical.

\end{acknowledgements}

\appendix

\section{New frequencies and spectroscopic parameters for CCD} \label{sec:ccd}

\cite{Yoshida2019} pointed out that the frequencies of some of the hyperfine components of the $N$\,=\,1-0 transition of CCD, as observed in L1527, are significantly different from those reported in the {\small CDMS} catalogue \citep{Muller2005}. The predictions from the {\small CDMS} catalogue are based in the experimental data reported by \cite{Bogey1985} and \cite{Vrtilek1985}, who observed the $N$\,=\,2-1, 3-2, 4-3, and 5-4 transitions, but only one hyperfine component of the $N$\,=\,1-0 transition. \citet{Yoshida2019} reported observed frequencies in L1527 for seven hyperfine components of the $N$\,=\,1-0 transition, with discrepancies with respect to the {\small CDMS} ones of up to 224\,kHz. The inaccuracy of the $N$\,=\,1-0 frequencies reported in the {\small CDMS} catalogue is due to the fact that they are calculated based on the laboratory measurements of higher $N$ transitions with a limited spectral resolution, which makes it difficult to discern between close hyperfine components.

In the analysis of our IRAM\,30m data of L483 we also noticed that the observed frequencies of the hyperfine components of the $N$\,=\,1-0 and 2-1 transitions of CCD differ from those reported in the {\small CDMS} catalogue. In order to provide more accurate predictions of the rotational spectrum of CCD we carried out a fit in which we only include the astronomical frequencies for the $N$\,=\,1-0 and 2-1 transitions and another fit in which we also include the laboratory frequencies for the $N$\,=\,3-2, 4-3, and 5-4 transitions. The spectroscopic parameters resulting from our fits are given in Table~\ref{table:ccd_constants}, where they are compared with the original values reported by \citet{Bogey1985} and \citet{Vrtilek1985}. The frequencies used in the fits are given in Table~\ref{table:ccd_frequencies}.

The spectroscopic parameters derived in our two fits are very similar. The main difference is found in the distortion constant $D$, which is affected by the inclusion of higher $N$ transitions when moving from the astronomical fit to the astronomical\,+\,laboratory fit. There is a notable change in the new value of $B$ derived in our fits compared to those derived by \citet{Bogey1985} and \citet{Vrtilek1985}, while the value of $D$ obtained in our combined fit agrees with those reported by \citet{Bogey1985} and \citet{Vrtilek1985}. The new value for $\gamma$ is determined more accurately and also agrees with those reported before. The larger differences are found for the hyperfine constants $b_F$ and $c$, whose previous values also differ between the two laboratory studies. The high resolution of our measurements is reflected in a higher accuracy determination for these hyperfine constants. The nuclear quadrupole coupling constant $eQq$ is also determined more precisely than before, but the previous values agree with the new one. We recommend the spectroscopic parameters derived in our combined, astronomical\,+\,laboratory, fit to predict the rotational spectrum of CCD.

\begin{table*}[hb!]
\small
\caption{Spectroscopic parameters of CCD (all in MHz).}
\label{table:ccd_constants}
\centering
\begin{tabular}{lcccc}
\hline \hline
Parameter             & Astronomical              & Astronomical + Laboratory           &	  \cite{Vrtilek1985}    &   \cite{Bogey1985}  \\
\hline
$B$                  &   36068.01625(297)    &   36068.02145(153)   &  36068.035(14)    &36068.0310(96)       \\
$D$                  &       0.06709(40)     &      0.067892(126)   &     0.0687(7)     &   0.06764(40)      \\
$\gamma$             &     $-$55.8334(34)    &      $-$55.8349(32)  &    $-$55.84(3)    &  $-$55.880(46)      \\
$b_F$$^{\rm(D)}$     &      6.9918(51)       &        6.9912(49)    &     6.35(7)       &   7.159(85)         \\
$c$$^{\rm(D)}$       &      1.8893(133)      &        1.8898(127)   &      1.59(26)     &    0.712(73)\,$^a$   \\
$eQq$$^{\rm(D)}$     &      0.2145(142)      &        0.2166(134)   &      0.21(9)      &    0.23(11)          \\
\hline
rms(kHz)             &        14.99		       &   			31.61			    & 			  --       &     --  \\
\hline
\end{tabular}
\tablenoteb{$^a$\,The parameter $c$ is denoted as $t$ in \cite{Bogey1985}.}
\end{table*}

\begin{table}
\small
\caption{Transition frequencies of CCD in MHz.}
\label{table:ccd_frequencies}
\centering
\begin{tabular}{{c@{\hspace{0.2cm}}c@{\hspace{0.2cm}}cc@{\hspace{0.2cm}}c@{\hspace{0.2cm}}crrl}}
\hline \hline
$N'$ & $J'$ &  $F'$ &  $N''$ & $J''$ & $F''$ & \multicolumn{1}{c}{$\nu_{obs}$} & \multicolumn{1}{c}{$\nu_{o}-\nu_{c}$} & \multicolumn{1}{c}{Ref.} \\
\hline
1 & 2 & 1 & 0 & 1 & 2 &  72098.573 $\pm$ 0.010 &  0.009 & (1) \\
1 & 2 & 2 & 0 & 1 & 2 &  72101.780 $\pm$ 0.015 & $-$0.025 & (2) \\
1 & 2 & 3 & 0 & 1 & 2 &  72107.716 $\pm$ 0.010 & $-$0.001 & (1) \\
1 & 2 & 1 & 0 & 1 & 1 &  72109.017 $\pm$ 0.015 & $-$0.033 & (2) \\
1 & 2 & 2 & 0 & 1 & 1 &  72112.290 $\pm$ 0.010 & $-$0.002 & (1) \\
1 & 1 & 2 & 0 & 1 & 2 &  72187.707 $\pm$ 0.010 & $-$0.008 & (1) \\
1 & 1 & 1 & 0 & 1 & 2 &  72189.711 $\pm$ 0.010 & $-$0.014 & (1) \\
1 & 1 & 2 & 0 & 1 & 1 &  72198.190 $\pm$ 0.010 & $-$0.011 & (1) \\
1 & 1 & 1 & 0 & 1 & 1 &  72200.208 $\pm$ 0.010 & $-$0.004 & (1) \\
2 & 3 & 3 & 1 & 2 & 3 & 144237.146 $\pm$ 0.010 &  0.001 & (1) \\
2 & 3 & 2 & 1 & 2 & 2 & 144239.737 $\pm$ 0.010 &  0.003 & (1) \\
2 & 3 & 4 & 1 & 2 & 3 & 144241.937 $\pm$ 0.010 & $-$0.000 & (1) \\
2 & 3 & 2 & 1 & 2 & 1 & 144243.028 $\pm$ 0.020 &  0.052 & (1) \\
2 & 3 & 3 & 1 & 2 & 2 & 144243.028 $\pm$ 0.020 & $-$0.027 & (1) \\
2 & 2 & 3 & 1 & 1 & 2 & 144296.719 $\pm$ 0.010 &  0.007 & (1) \\
2 & 2 & 2 & 1 & 1 & 1 & 144297.544 $\pm$ 0.010 &  0.008 & (1) \\
2 & 2 & 1 & 1 & 1 & 1 & 144299.042 $\pm$ 0.010 &  0.014 & (1) \\
2 & 2 & 2 & 1 & 1 & 2 & 144299.548 $\pm$ 0.010 &  0.002 & (1) \\
2 & 2 & 1 & 1 & 1 & 2 & 144301.017 $\pm$ 0.020 & $-$0.020 & (1) \\
2 & 2 & 3 & 1 & 2 & 3 & 144376.713 $\pm$ 0.010 &  0.003 & (1) \\
2 & 2 & 3 & 1 & 2 & 2 & 144382.614 $\pm$ 0.010 & $-$0.007 & (1) \\
2 & 2 & 1 & 1 & 2 & 2 & 144386.909 $\pm$ 0.020 & $-$0.038 & (1) \\
2 & 2 & 2 & 1 & 2 & 1 & 144388.688 $\pm$ 0.010 & $-$0.009 & (1) \\
2 & 2 & 1 & 1 & 2 & 1 & 144390.221 $\pm$ 0.010 &  0.031 & (1) \\
3 & 4 & 4 & 2 & 3 & 4 & 216368.560 $\pm$ 0.050 &  0.020 & (3) \\
3 & 4 & 3 & 2 & 3 & 3 & 216369.990 $\pm$ 0.070 & $-$0.010 & (3) \\
3 & 4 & 5 & 2 & 3 & 4 & 216372.830 $\pm$ 0.020 & $-$0.016 & (3) \\
3 & 4 & 3 & 2 & 3 & 2 & 216373.320 $\pm$ 0.020 & $-$0.003 & (3) \\
3 & 4 & 4 & 2 & 3 & 3 & 216373.320 $\pm$ 0.020 & $-$0.012 & (3) \\
3 & 3 & 4 & 2 & 2 & 3 & 216428.320 $\pm$ 0.020 &  0.092 & (3) \\
3 & 3 & 3 & 2 & 2 & 2 & 216428.320 $\pm$ 0.020 & $-$0.086 & (3) \\
3 & 3 & 2 & 2 & 2 & 1 & 216428.760 $\pm$ 0.040 & $-$0.090 & (3) \\
3 & 3 & 2 & 2 & 2 & 2 & 216430.340 $\pm$ 0.060 & $-$0.002 & (3) \\
3 & 3 & 3 & 2 & 2 & 3 & 216431.260 $\pm$ 0.050 &  0.020 & (3) \\
4 & 5 & 6 & 3 & 4 & 5 & 288499.000 $\pm$ 0.050 &  0.147 & (4) \\
4 & 5 & 5 & 3 & 4 & 4 & 288499.000 $\pm$ 0.050 & $-$0.123 & (4) \\
4 & 5 & 4 & 3 & 4 & 3 & 288499.000 $\pm$ 0.050 & $-$0.126 & (4) \\
4 & 4 & 5 & 3 & 3 & 4 & 288554.590 $\pm$ 0.050 &  0.151 & (4) \\
4 & 4 & 4 & 3 & 3 & 3 & 288554.590 $\pm$ 0.050 &  0.075 & (4) \\
4 & 4 & 3 & 3 & 3 & 2 & 288554.590 $\pm$ 0.050 & $-$0.177 & (4) \\
5 & 6 & 7 & 4 & 5 & 6 & 360618.340 $\pm$ 0.150 &  0.003 & (4) \\
5 & 6 & 6 & 4 & 5 & 5 & 360618.340 $\pm$ 0.150 & $-$0.168 & (4) \\
5 & 6 & 5 & 4 & 5 & 4 & 360618.340 $\pm$ 0.150 & $-$0.173 & (4) \\
5 & 5 & 6 & 4 & 4 & 5 & 360674.170 $\pm$ 0.150 &  0.155 & (4)  \\
5 & 5 & 5 & 4 & 4 & 4 & 360674.170 $\pm$ 0.150 &  0.114 & (4) \\
5 & 5 & 4 & 4 & 4 & 3 & 360674.170 $\pm$ 0.150 & $-$0.048 & (4) \\
\hline
\end{tabular}
\tablenotea{References: (1) Frequencies derived in this work from Gaussian fits to the lines observed in L483, adopting $V_{\rm LSR}$ = 5.30 km s$^{-1}$ \citep{Agundez2019}. (2) These two lines are affected by a negative frequency-switching artifact in our L483 data and thus frequencies are taken from the L1527 data of \cite{Yoshida2019}. (3) Laboratory data of \cite{Vrtilek1985}. (4) Laboratory data of \cite{Bogey1985}.}
\end{table}

\end{document}